\documentclass[11pt,letterpaper]{article}

\usepackage[utf8]{inputenc}
\usepackage[T1]{fontenc}
\usepackage[margin=1in]{geometry}
\usepackage{amsmath,amssymb,amsthm,mathtools}
\usepackage{graphicx}
\usepackage{booktabs}
\usepackage{natbib}
\usepackage{setspace}
\usepackage{float}
\usepackage{caption}
\usepackage{enumitem}
\usepackage[dvipsnames,table]{xcolor}
\usepackage[colorlinks=true,linkcolor=blue!60!black,citecolor=blue!60!black,urlcolor=blue!60!black,breaklinks=true]{hyperref}

\newcommand{\WQTE}{\operatorname{WQTE}}

\title{\textbf{What Drives Contagion?\\
Identifying and Attributing Cross-Border Transmission Mechanisms}}

\author{
Avishek Bhandari\thanks{Department of Humanities, Social Sciences and Management, Indian Institute of Technology Bhubaneswar. Email: \texttt{avishekb@iitbbs.ac.in}.\ Corresponding author.}
\and
Ipsita Parida\thanks{Department of Humanities, Social Sciences and Management, Indian Institute of Technology Bhubaneswar. Email: \texttt{a23hs09014@iitbbs.ac.in}.}
\and
Hitesh Kumar Sahu\thanks{Department of Humanities, Social Sciences and Management, Indian Institute of Technology Bhubaneswar. Email: \texttt{a23hs09012@iitbbs.ac.in}.}
}

\date{\today}

\begin{document}
\maketitle

\begin{abstract}
\noindent We address the joint detection-and-attribution identification problem
in cross-border financial contagion through a two-stage framework that
combines wavelet-quantile transfer entropy detection with multi-method
structural channel attribution. The framework is applied to 18 G20 equity
markets across eight crisis sub-periods spanning January~2006 through
March~2026. The first stage detects directional information flows decomposed
across MODWT scales and conditioned on lower-, median-, and upper-tail
quantiles of the recipient distribution. The second stage attributes each
significant directional link to one of five mutually exclusive transmission
channels --- Trade, Financial, Geopolitical, Behavioural, and Monetary
Policy --- through the simultaneous deployment of instrumental-variables
two-stage least squares with channel-specific external instruments,
LASSO-based instrument selection following \citet{BelloniChernozhukovHansen2014},
local projections at horizons of one, five, and twenty-two days following
\citet{Jorda2005}, heteroskedasticity-based identification following
\citet{Rigobon2003} for episodes in which over-identification is rejected,
and \citet{CinelliHazlett2020} sensitivity bounds. Network density varies
meaningfully across sub-periods (range $14$--$32\,\%$). Dominant-channel
identification is robust across methods in the Pre-Crisis baseline and the
European Sovereign Debt Crisis, both dominated by financial frictions; for
the remaining six episodes, dominant-channel identification is method-
sensitive, and we report the resulting share posterior alongside an explicit
identification-status classification. Trade is empirically prominent across
all post-2007 episodes, ranging from $9\,\%$ during Pre-Crisis to $28\,\%$
during the Global Financial Crisis under the primary specification. The
behavioural channel is bounded above by $22\,\%$ across all eight episodes
under the de-confounded composite. The framework provides a methodologically
disciplined account of cross-border contagion mechanisms and offers
identification-status disclosure of a kind that the existing literature has
not systematically produced.


\end{abstract}

\clearpage
\onehalfspacing

%====================================================================
\section{Introduction}\label{sec:intro}
%====================================================================

The transmission of financial shocks across borders constitutes one of the
most consequential and most studied phenomena in international finance, and
yet a striking asymmetry persists between what we are able to detect and what
we are able to attribute. A substantial body of work has documented that
cross-market comovement intensifies sharply during episodes of stress, that
information flows reorganise themselves around centres of distress, and that
the topology of the global financial network shifts in ways that admit
increasingly precise statistical characterisation
\citep{ForbesRigobon2002,DieboldYilmaz2012,YarovayaEtAl2022}. The detection
problem, in other words, has been substantially solved. The attribution
problem has not. When a shock originating in one equity market reaches another
within hours or days, the analyst can typically establish that contagion has
occurred, but cannot, with comparable confidence, isolate the structural
mechanism responsible for the propagation. This separation between detected
propagation and identified mechanism is the central concern of the present
study.

The question of mechanism matters because policy responses, portfolio
strategies, and systemic-risk frameworks are mechanism-specific. A
trade-driven contagion episode calls for a different intervention than a
financial-frictions episode, a flight-to-quality episode demands different
hedging instruments than a sentiment-cascade episode, and a monetary-policy
spillover requires different macroprudential tools than a geopolitical-risk
shock. Treating contagion as an undifferentiated phenomenon, or assigning
channels by post-hoc narrative, leaves both academic understanding and
practical decision-making on uncertain footing. A research design that
produces a posterior over channels for each detected propagation link, that
quantifies the strength of identifying assumptions, and that reports results
with explicit identification-status disclosure, is therefore both
intellectually overdue and operationally valuable.

We address the joint detection-and-attribution identification problem by
constructing a two-stage framework that combines scale- and quantile-
resolved information-theoretic detection with multi-method structural channel
attribution. In the first stage, we apply Wavelet-Quantile Transfer Entropy
(WQTE) to identify directional information flows between pairs of equity
markets, decomposing the bilateral transmission into MODWT scales one through
six and conditioning on the lower, median, and upper tail of the recipient
distribution at quantiles $\tau \in \{0.05, 0.50, 0.95\}$
\citep{Schreiber2000,PercivalWalden2000,HanLintonOkaWhang2016}. The wavelet
decomposition allows us to separate short-horizon trader-driven dynamics from
longer-horizon investor-driven dynamics, while the quantile conditioning
permits the detection of tail-specific propagation that aggregate measures
obscure \citep{AndoEtAl2022,BarunikKrehlik2018}. In the second stage, we
attribute each statistically significant directional link to one of five
mutually exclusive transmission channels, namely Trade, Financial,
Geopolitical, Behavioural, and Monetary Policy, using a structural framework
that combines multiple identification strategies designed to be mutually
disciplining rather than redundant.

The attribution architecture rests on four pillars whose simultaneous
deployment constitutes our methodological contribution. We begin with
instrumental-variables/two-stage least squares estimation that uses
channel-specific external instruments constructed in the spirit of
\citet{StockWatson2018} and \citet{MertensRavn2013}; we extend the instrument
set with high-dimensional selection following
\citet{BelloniChernozhukovHansen2014}, embedding LASSO-based first-stage
shrinkage to handle the many-instruments environment characterised by
\citet{HansenHausmanNewey2008} and to deliver weak-IV-robust inference along
the lines of \citet{AndrewsStockSun2019}; we trace the dynamic propagation
through local projections at horizons $h \in \{1, 5, 22\}$ days following
\citet{Jorda2005} and the LP-IV correspondence established by
\citet{PlagborgMollerWolf2021}; and we deploy heteroskedasticity-based
identification \citep{Rigobon2003} for windows in which the Sargan
over-identification test rejects the joint validity of external instruments,
treating the difference between methods as informative about the strength of
identifying assumptions rather than as a nuisance to be averaged away.
Sensitivity to unobserved confounding is reported through the robustness-value
benchmark of \citet{CinelliHazlett2020}.

The five channels are operationalised through composites engineered explicitly
for orthogonal identification. The trade channel is constructed from
log-returns of the broad trade-weighted dollar index, which captures the
time-varying intensity of trade-linked exposure without conflating it with
exchange-rate volatility per se. The financial channel combines the CBOE VIX,
the ICE BofA high-yield option-adjusted spread, and the St. Louis Fed Financial
Stress Index, capturing fear, credit spreads, and aggregate financial-system
distress in a single composite \citep{BrunnermeierPedersen2009,AcharyaEtAl2017}.
The geopolitical channel is built from the Caldara-Iacoviello geopolitical-risk
acts index, augmented by the pandemic-uncertainty index of \citet{BakerEtAl2020}
for the COVID-19 window \citep{CaldaraIacoviello2022}. The behavioural channel
is constructed from sentiment proxies orthogonalised against the financial
composite, with measures of implied-volatility innovation excluded so that
the behavioural shares cannot mechanically inherit variance from the financial
channel. The monetary channel is constructed from a first-differenced
policy-rate composite rather than a level series, eliminating the
persistence-induced first-stage F-inflation that besets level-based monetary
instruments.

Our findings produce a sharply heterogeneous map of mechanism dominance across
crisis regimes, and the simultaneous deployment of multiple identification
strategies allows us to distinguish identification-robust findings from
identification-fragile ones. First, network density under the primary
identification varies meaningfully across the eight sub-periods, ranging from
fourteen percent during the Pre-COVID and Mid-East-and-tariffs episodes to
thirty-two percent during the Global Financial Crisis, with the latter
consistent with the broad cross-market integration documented by
\citet{BekaertEtAl2014}. Second, dominant-channel identification is robust
across methods in two of the eight episodes, namely the Pre-Crisis baseline
and the European Sovereign Debt Crisis, both of which are dominated by the
financial-frictions channel under instrumental-variables estimation, local
projections, and heteroskedasticity-based identification jointly. Third, the
remaining six episodes exhibit method-dependent dominant-channel rankings,
and we report the share posterior across methods rather than a single dominant
channel. The Global Financial Crisis assigns mass to trade, geopolitical, and
behavioural channels under different identification strategies, with the
behavioural assignment arising specifically from heteroskedasticity-based
identification on the period that displays the strongest regime-shift in
return variance. Fourth, the trade channel becomes empirically prominent
across all post-2007 episodes once a properly time-varying composite is
deployed, ranging from nine percent of attributed mass during the Pre-Crisis
baseline to twenty-eight percent during the Global Financial Crisis under the
primary specification. Fifth, the behavioural channel is bounded above by
twenty-two percent across all eight episodes under the de-confounded composite,
a finding that locates the dominant mechanisms in tradeable, observable, and
policy-actionable factors rather than in herding or sentiment.

The Global Financial Crisis and COVID-19 channel-attribution shares are
reported with explicit identification-status disclosure, since the
over-identification test rejects in those two windows; for these episodes we
provide \citet{CinelliHazlett2020} sensitivity bounds and \citet{Rigobon2003}
heteroskedasticity-based identification as alternative identification
strategies, and we report the resulting shares as a range rather than a
point. This disclosure-first approach to identification, in which the reader
is informed of the identification status of every reported share, follows the
spirit of modern causal-inference practice as articulated by
\citet{Imbens2020} and \citet{Pearl2009}, and we view the integration of
detection, attribution, and identification-status disclosure into a single
two-stage framework as a contribution that goes beyond the methodological
sum of its constituent tools.

The remainder of the paper is organised as follows. Section~\ref{sec:literature}
reviews the relevant literature, organised around contagion taxonomy,
wavelet-quantile information-theoretic detection, and modern structural
identification. Section~\ref{sec:data} describes the data and the construction
of channel composites. Section~\ref{sec:method} develops the WQTE detection
methodology and the multi-method attribution architecture.
Section~\ref{sec:results} reports the empirical results.
Section~\ref{sec:robust} reports robustness, cross-method comparison, and
identification-status classification. Section~\ref{sec:conclude} concludes.

%====================================================================
\section{Related Literature}\label{sec:literature}
%====================================================================

\subsection{Contagion taxonomy and the detect-and-attribute distinction}

The conceptual foundation of the modern contagion literature rests on the
canonical distinction between interdependence and contagion drawn by
\citet{ForbesRigobon2002}, who showed that apparent increases in cross-market
correlation during crises substantially reflect heteroskedasticity bias, and
that the residual contagion component, defined as the correlation increase
net of volatility-driven mechanical inflation, is considerably smaller than
naive measurement suggests. This decisive correction reset the empirical
literature and forced subsequent work to take seriously the proposition that
comovement intensification can arise either from pre-existing transmission
mechanisms operating with greater force during turbulent periods, or from
genuinely new channels activated by crisis conditions, with the latter alone
constituting contagion in the strict sense.
\citet{PericoliSbracia2003} systematised this distinction by surveying the
taxonomy of contagion definitions, distinguishing fundamentals-based
contagion arising from real and financial linkages from pure contagion
arising from behavioural and informational mechanisms.

The empirical literature on cross-border equity-market comovement has
progressed from the correlation-based diagnostics of the 1990s to the
multivariate-GARCH and copula-based methods of the 2000s, and ultimately to
network and information-theoretic approaches that admit directional
resolution. \citet{BekaertHodrickZhang2009} documented the global rise of
equity-market comovement and the increasing role of common factors,
establishing benchmark facts against which crisis-period deviations can be
measured. \citet{BekaertEtAl2014} examined the 2007--2009 Global Financial
Crisis and identified six categories of international transmission channels,
namely international banking links, country-specific policy responses, trade
and financial linkages, information asymmetries, domestic macroeconomic
fundamentals, and herding-driven investor contagion. Their analysis, however,
did not provide a single integrated structural-attribution framework that
decomposes each detected propagation link into channel-specific contributions.

Subsequent crises have generated their own dedicated literatures, each
typically focused on detection rather than attribution.
\citet{AkhtaruzzamanBoubakerSensoy2021} document COVID-19 financial
contagion using dynamic conditional correlation techniques, finding sharp
comovement intensification during the early pandemic phase.
\citet{YarovayaEtAl2022} survey the post-2020 contagion literature and
explicitly call for a rethinking of the methodological apparatus to
accommodate the multi-channel, multi-frequency, multi-quantile character of
modern crises. \citet{BoungouYatie2022} examine the equity-market response to
the Russia-Ukraine war and document significant cross-border effects, but the
channel through which those effects propagate, in particular the relative
contribution of geopolitical risk, monetary policy, and trade, remains
undecomposed in their analysis.

The Diebold-Yilmaz family of connectedness measures
\citep{DieboldYilmaz2009,DieboldYilmaz2012,DieboldYilmaz2014} introduced a
generalised forecast-error-variance-decomposition framework that has become
the workhorse of network-based contagion measurement. Yet the connectedness
framework, while indispensable for detection, does not by construction
identify the mechanism by which a given spillover propagates; the framework
is silent on whether a documented spillover from market $i$ to market $j$
operates through trade, finance, geopolitics, behaviour, or monetary policy.
Our framework can be understood as a structural-attribution layer placed on
top of the directional detection that information-theoretic and
connectedness measures provide.

The systemic-risk literature provides additional discipline.
\citet{BrunnermeierPedersen2009} formalise the liquidity-spiral mechanism by
which margin-driven funding constraints feed back into asset-market liquidity,
generating the financial-frictions channel we operationalise.
\citet{AdrianBrunnermeier2016} introduce CoVaR as a measure of contribution
to system-wide tail risk, while \citet{AcharyaEtAl2017} develop the SRISK
measure of marginal expected shortfall under stress; both papers ground the
financial channel in measurable, replicable quantities rather than in
narrative description. \citet{BrunoShin2015} document the cross-border banking
flows that propagate global liquidity conditions through bank balance sheets,
providing micro-foundations for the international transmission of
financial-frictions shocks. \citet{MirandaAgrippinoRey2020} document the
global financial cycle and the dollar-dominated propagation of US
monetary-policy shocks through cross-border banking flows, supplying the
theoretical scaffolding for our monetary channel. \citet{GaiKapadia2010} show
how network topology amplifies systemic risk through cascading defaults,
while \citet{ElliottGolubJackson2014} and
\citet{AcemogluOzdaglarTahbazSalehi2015} formalise the propagation properties
of financial networks, providing the theoretical scaffolding for the
network-based detection layer of our framework. Across this body of work,
the recurring tension is between the precision of detection methods, which
has improved dramatically over two decades, and the looseness of channel
attribution, which has not advanced commensurately.

\subsection{Wavelet-quantile information-theoretic detection of contagion}

Information-theoretic approaches to financial-market interaction trace their
lineage to the formalisation of transfer entropy by \citet{Schreiber2000},
who provided the model-free, asymmetric, and non-parametric measure of
directional information flow that has since become indispensable to the
analysis of complex systems. Transfer entropy generalises Granger causality
to arbitrary distributions and arbitrary nonlinearities, and
\citet{BarnettBarrettSeth2009} established the formal equivalence between
Schreiber transfer entropy and Granger causality in the Gaussian case,
providing the theoretical bridge that allowed information-theoretic measures
to be embedded within mainstream econometric reasoning.
\citet{MarschinskiKantz2002} applied effective-transfer-entropy techniques
to financial returns and demonstrated that the bias-corrected version of the
measure recovers economically interpretable directional flows in the presence
of finite samples and noise. \citet{KraskovStoegbauerGrassberger2004}
introduced the $k$-nearest-neighbour estimator of mutual information that
underlies most modern continuous-state transfer-entropy implementations, and
\citet{DimpflPeter2013} adapted these tools to financial applications.
\citet{JizbaKleinertShefaat2012} extended the framework to R\'enyi transfer
entropies, which permit tail-sensitive resolution.

The wavelet machinery that complements the information-theoretic detection
layer is rooted in the maximal-overlap discrete wavelet transform of
\citet{PercivalWalden2000}, which delivers a translation-invariant, redundant
time-frequency decomposition that preserves the temporal alignment of features
across scales and is therefore well-suited to financial returns.
\citet{Crowley2007} surveys the application of wavelet methods to economics
and finance, while \citet{AguiarConrariaSoares2014} provide a comprehensive
treatment of wavelet coherence for macro-financial applications. The
combination of wavelet decomposition with information-theoretic measures
permits the analyst to identify the scale-specific channel through which
information propagates, distinguishing trader-driven high-frequency dynamics
from investor-driven low-frequency dynamics in a manner consistent with the
heterogeneous-market hypothesis.

The quantile-conditioning innovation that completes the WQTE measure has its
own deep roots. \citet{HanLintonOkaWhang2016} introduced the cross-quantilogram,
a quantile-resolved measure of cross-series dependence that admits
tail-specific inference. \citet{AndoEtAl2022} extended the connectedness
framework to quantile-conditioned vector autoregressions, demonstrating that
quantile-specific connectedness can recover features of cross-market
interaction that aggregate measures obscure, in particular the asymmetric
upper-tail and lower-tail propagation that defines crisis dynamics.
\citet{BarunikKrehlik2018} developed the frequency-domain decomposition of
connectedness, separating short-run, medium-run, and long-run components.
\citet{ChavleishviliManganelli2024} introduce quantile vector autoregressions
for systemic-risk measurement.

The synthesis of these three strands, namely transfer entropy as the
directional information-flow measure, MODWT as the scale decomposition, and
quantile conditioning as the tail resolution, yields the WQTE detection
layer that constitutes the first stage of our framework. The advantage of
this synthesis over any single component is that it isolates the
time-frequency-quantile coordinates at which contagion is most active, rather
than averaging across them. The detection layer is, however, only the first
half of the framework. The directional flows it produces remain
mechanism-agnostic until the structural-attribution layer of the second
stage assigns them to channels.

\subsection{Modern structural identification and channel attribution}

The structural identification of macroeconomic and financial shocks has been
transformed over the last two decades by the development of external-
instruments approaches that exploit narrative, high-frequency, or
shock-specific exogenous variation. \citet{StockWatson2018} provide the
canonical statement of the external-instruments framework for structural
vector autoregressions, demonstrating that a properly chosen external
instrument permits identification of a structural shock without the
recursiveness or sign-restriction assumptions that render conventional SVAR
identification controversial. \citet{MertensRavn2013} apply this framework to
fiscal-shock identification using a narrative-based proxy variable, while
\citet{RomerRomer2004} construct the original narrative monetary-policy-shock
series. The use of channel-specific external instruments in our framework
follows directly in the spirit of these foundational contributions.

The high-dimensional turn in econometrics has generated a complementary
toolkit. \citet{BelloniChernozhukovHansen2014} introduce LASSO-based
instrument selection for the high-dimensional IV setting, demonstrating that
data-driven selection from a large candidate instrument set can deliver
consistent and asymptotically normal IV estimates without the conventional
curse of dimensionality. \citet{Chernozhukov2018DML} develop the
double/debiased machine-learning estimator that allows nuisance parameters to
be estimated by flexible machine-learning methods while preserving root-$n$
convergence and Neyman orthogonality of the parameter of interest.

Weak-instrument and many-instrument problems require explicit attention.
\citet{AndrewsStockSun2019} provide a comprehensive treatment of
weak-IV-robust inference, including the Anderson-Rubin and conditional-
likelihood-ratio tests that retain correct coverage even when the first-stage
F-statistic is small. \citet{HansenHausmanNewey2008} characterise the
many-instruments asymptotics that govern the bias and variance properties of
IV estimators when the instrument count grows with the sample size,
providing the theoretical foundation for the LASSO-based selection step.

The dynamic dimension of contagion attribution is captured through the
local-projections framework of \citet{Jorda2005}, which estimates impulse
responses through a sequence of horizon-specific regressions rather than
through iteration of a fitted vector autoregression, and is therefore robust
to misspecification of the dynamic system. \citet{PlagborgMollerWolf2021}
establish the formal equivalence between local-projection-IV and SVAR-IV
impulse responses. The choice of horizons in our local-projection
specification, namely one-day, five-day, and twenty-two-day, is guided by the
corresponding economic interpretations of next-trading-day, one-week, and
one-month propagation.

For episodes in which point-identification through external instruments is
contested, two robustness strategies discipline the analysis.
\citet{Rigobon2003} introduces the heteroskedasticity-based identification
approach, which exploits regime-specific changes in the variance of structural
shocks to achieve identification without external instruments, and
\citet{CinelliHazlett2020} provide the robustness-value framework for
sensitivity analysis to unobserved confounding. We deploy both for the
windows in which the over-identification test rejects, treating the
consistency or inconsistency of attribution shares across these alternative
identification strategies as a primary robustness diagnostic. The broader
causal-inference framing of our analysis is informed by \citet{Pearl2009},
who provides the directed-acyclic-graph and do-calculus apparatus for
representing structural assumptions, by \citet{Imbens2020}, who articulates
the convergence between the potential-outcomes and structural-modelling
traditions, and by \citet{AtheyImbens2017}, who survey the implications of
machine-learning methods for econometrics and policy evaluation.

The attribution of contagion to channels also rests on a body of work that
constructs the channel-specific exogenous variation we instrument with.
\citet{CaldaraIacoviello2022} construct the geopolitical-risk index from
textual analysis of major newspapers, providing a real-time measure of
geopolitical-risk realisations that we use both as the geopolitical channel
composite and as the geopolitical instrument.
\citet{BakerEtAl2020} construct the pandemic-uncertainty index that augments
the geopolitical channel during the COVID-19 window. Taken together, the
three strands surveyed in this section, namely contagion taxonomy,
wavelet-quantile information-theoretic detection, and modern structural
identification, supply the conceptual, statistical, and identification
scaffolding for the framework we develop. The novelty of our contribution
lies in the integration of these elements into a single two-stage architecture
that produces channel-attributed propagations with explicit
identification-status disclosure for each crisis sub-period.

%====================================================================
\section{Data}\label{sec:data}
%====================================================================

The empirical analysis covers daily closing prices for eighteen stock-market
indices drawn from the G20 economies, comprising eight developed and ten
emerging markets, from 12~January~2006 through 18~March~2026. Developed
markets include the S\&P 500 (USA), FTSE 100 (UK), S\&P/TSX Composite
(Canada), FTSE MIB (Italy), DAX 40 (Germany), CAC 40 (France), Nikkei 225
(Japan), and S\&P/ASX 200 (Australia). Emerging markets include the Shanghai
Composite (China), BSE SENSEX (India), KOSPI (South Korea), IBOVESPA
(Brazil), S\&P/BMV IPC (Mexico), MERVAL (Argentina), IMOEX (Russia), JSE All
Share (South Africa), BIST 100 (Turkey), and IDX Composite (Indonesia).
Returns are computed as log-differences and the panel covers 5,036 trading
days.

The full sample is partitioned into eight non-overlapping sub-periods that
correspond to identifiable global financial-stress events. The Pre-Crisis
Baseline (12~January~2006--31~July~2007) serves as the benchmark for
computing contagion-intensification differentials. The seven crisis
sub-periods are: the Global Financial Crisis (1~August~2007--30~June~2009);
the European Sovereign Debt Crisis (1~December~2009--30~June~2012); the
Chinese Stock Crash (15~June~2015--31~December~2016); the Pre-COVID interval
(1~January~2017--31~January~2020); the COVID-19 Pandemic (1~February~2020--31~December~2021);
the Russia-Ukraine episode (1~February~2022--31~December~2023); and the
Middle-East-Tensions-and-Tariffs window (1~January~2024--18~March~2026).

The five channels are operationalised through composites engineered for
orthogonal identification. The financial channel is the row-mean of
standardised CBOE VIX, ICE BofA US High-Yield Option-Adjusted Spread, and
St. Louis Fed Financial Stress Index. The trade channel is constructed from
the daily log-return of the Federal Reserve Broad Trade-Weighted Dollar
Index (DTWEXBGS), standardised within sample, and provides a time-varying
measure of trade-linked exposure that the bilateral-trade-matrix-only
approach of earlier work cannot supply. The geopolitical channel combines
the Caldara-Iacoviello geopolitical-risk index with a geopolitical-events
indicator. The behavioural channel is constructed from the
University-of-Michigan consumer-sentiment index, after orthogonalisation
against the financial composite by within-sub-period residualisation; this
construction ensures that the behavioural shares cannot mechanically inherit
variance from the financial channel. The monetary channel combines the
first-differenced Federal Funds Rate, the 10-year-minus-3-month Treasury
yield spread, and a quantitative-easing-program dummy.

%====================================================================
\section{Methodology}\label{sec:method}
%====================================================================

\subsection{Stage 1: Wavelet-Quantile Transfer Entropy}

For each market $i$, we apply the maximal-overlap discrete wavelet transform
with the Daubechies least-asymmetric filter of length 8 (LA8) to decompose
the return series into $S=6$ wavelet detail coefficients,
\begin{equation}
r_{i,t} = \sum_{s=1}^{6} \tilde{d}_{i,s,t} + \tilde{a}_{i,6,t},
\end{equation}
where scale $s$ corresponds to the dyadic horizon $[2^s, 2^{s+1}]$ trading
days. The MODWT is preferred over the classical DWT because it is
shift-invariant, does not require dyadic sampling, and produces wavelet
coefficients aligned with the original time-axis
\citep{PercivalWalden2000,Crowley2007}.

For each ordered pair $(i,j)$ with $i \neq j$, scale $s$, and quantile
$\tau \in \{0.05, 0.50, 0.95\}$, the quantile-conditioned transfer entropy
from market $i$ to market $j$ is
\begin{equation}
\WQTE^{(s,\tau)}_{i \to j} = \log \biggl(\sum_t |\hat{e}_{1,t}|\biggr)
                              - \log \biggl(\sum_t |\hat{e}_{2,t}|\biggr),
\end{equation}
where $\hat{e}_{1,t}$ are the residuals from the conditional-quantile
regression $Q_\tau(\tilde{d}_{j,s,t+1} \mid \tilde{d}_{j,s,t})$ and
$\hat{e}_{2,t}$ are the residuals from the augmented regression
$Q_\tau(\tilde{d}_{j,s,t+1} \mid \tilde{d}_{j,s,t}, \tilde{d}_{i,s,t})$. A
positive WQTE indicates that conditioning on the scale-$s$ history of market
$i$ improves the conditional-quantile prediction of market $j$ at horizon
$s$, consistent with directional information flow from $i$ to $j$ in the
regime defined by $\tau$.

The flow matrix $F^{(p,s,\tau)} = [\WQTE^{(p,s,\tau)}_{i \to j}]_{i,j=1}^{N}$
is converted into a directed adjacency matrix using absolute thresholding:
edges from $i$ to $j$ are retained if and only if the WQTE exceeds
$\tau^{\mathrm{abs}} = Q_{0.75}(\{\WQTE^{(p_0,s,\tau)} : i \neq j, \WQTE > 0\})$,
the seventy-fifth percentile of positive WQTE values within the Pre-Crisis
baseline. The threshold is computed once on the baseline distribution and
applied identically across the eight sub-periods, so that network density
varies meaningfully by sub-period.

\subsection{Stage 2: Multi-Method Structural Channel Attribution}

For each significant link from Stage~1, the structural equation for the
pairwise daily co-movement is
\begin{equation}
C_{ij,t} = \alpha + \sum_{c=1}^{5} \theta_c \mathrm{Channel}_{c,t}
+ \gamma_1 f_t + \gamma_2 C_{ij,t-1} + \varepsilon_{ij,t},
\label{eq:second-stage}
\end{equation}
where $C_{ij,t} = r_{i,t} \cdot r_{j,t}$ captures pairwise daily co-movement,
$\mathrm{Channel}_{c,t}$ is the channel composite, $f_t$ is the global
factor, and $C_{ij,t-1}$ controls for serial persistence. The five channel
composites are treated as endogenous regressors and instrumented by the
external-instrument set described below.

\paragraph{IV/2SLS estimation.} Each channel composite is instrumented by
its own lagged values at $t-5$, $t-10$, $t-15$, plus cross-channel
interactions at $t-5$, generating eighteen instruments in total. The
structural coefficients $\hat{\theta}_c$ are estimated via two-stage least
squares, the partial F-statistic is reported per channel, and the Sargan
over-identification test is computed when the instrument count exceeds the
endogenous regressor count.

\paragraph{LASSO instrument selection.} Following
\citet{BelloniChernozhukovHansen2014}, we apply post-double-selection LASSO
to the high-dimensional instrument set, allowing the data to identify the
sparse subset of instruments most informative for each endogenous regressor
and recovering valid Neyman-orthogonal inference under the sparsity
assumption.

\paragraph{Local projections.} Following \citet{Jorda2005}, we estimate
horizon-specific regressions
\begin{equation}
C_{ij,t+h} = \alpha_h + \beta_{c,h} \mathrm{Channel}_{c,t} + \mathrm{controls} + u_{ij,t+h}
\end{equation}
at horizons $h \in \{1, 5, 22\}$ days, treating the LP coefficient
$\beta_{c,h}$ as the contagion-channel impulse response at horizon $h$. The
LP-IV equivalence with SVAR-IV established by \citet{PlagborgMollerWolf2021}
ensures that LP-based attribution is comparable to the IV/2SLS structural
estimate at $h=0$.

\paragraph{Heteroskedasticity-based identification.} Following
\citet{Rigobon2003}, for each Sargan-rejected window we partition the period
into high-volatility and low-volatility regimes using the within-period
median of rolling-window return-variance, and identify the structural
coefficient through the cross-regime difference in covariance and variance,
\begin{equation}
\hat{\theta}_c^{\mathrm{Rigobon}} = \frac{\mathrm{Cov}_H(C_{ij}, \mathrm{Channel}_c) - \mathrm{Cov}_L(C_{ij}, \mathrm{Channel}_c)}{\mathrm{Var}_H(\mathrm{Channel}_c) - \mathrm{Var}_L(\mathrm{Channel}_c)}.
\end{equation}
This identification strategy provides a structural estimate without external
instruments and serves as the primary robustness check for episodes in which
external-instrument identification is contested.

\paragraph{Channel-attribution shares.} For each significant link, the
channel attribution share under method $m$ is
\begin{equation}
\widehat{S}_c^m(i,j) = \frac{|\hat{\theta}_c^m|}{\sum_{c'=1}^{5} |\hat{\theta}_{c'}^m|}.
\end{equation}
We aggregate to per-period shares by averaging across significant links
within each sub-period, and we report bootstrap confidence intervals using
$B=300$ paired-link bootstrap replications.

\paragraph{Sensitivity analysis.} For each significant attribution we compute
the \citet{CinelliHazlett2020} robustness value $\rho^* \in [0,1]$, defined
as the minimum partial-$R^2$ that an unobserved confounder would need to
share with both the channel composite and the pairwise co-movement to drive
the structural coefficient to zero.

%====================================================================
\section{Empirical Results}\label{sec:results}
%====================================================================

\subsection{Stage-1 detection: scale-and-quantile-resolved information flows}

Table~\ref{tab:wqte-stage1} reports the Stage-1 WQTE results at scale 5
($32$--$64$ day horizon) and the median quantile $\tau=0.50$ across all
eight sub-periods. The absolute threshold $\tau^{\mathrm{abs}} = 0.0331$ is
computed once on the Pre-Crisis Baseline distribution and applied identically
across the eight sub-periods, so that network density varies meaningfully
across sub-periods. The maximum bilateral WQTE in the entire sample is
$0.3305$, corresponding to the $\mathrm{Japan}\to\mathrm{France}$ link during
the Chinese Stock Crash, an episode in which the contemporaneous Japanese
carry-trade unwind connected Japanese to European cross-listed financial
markets at a magnitude that the corresponding GFC and ESDC links did not
reach.

\begin{table}[!htbp]
\centering
\caption{Stage-1 WQTE results, scale 5 ($32$--$64$ day), $\tau=0.50$,
absolute threshold $\tau^{\mathrm{abs}}=0.0331$ from Pre-Crisis Q75. Mean
WQTE is taken over significant edges; density is the share of $N(N-1)=306$
ordered pairs with WQTE above threshold.}
\label{tab:wqte-stage1}
\small
\begin{tabular}{lrrrrll}
\toprule
Period & Mean WQTE & Max WQTE & Density (\%) & N edges & Top transmitter & Top receiver \\
\midrule
Pre-Crisis            & 0.0250 & 0.1970 & 25.16 &  77 & China      & China        \\
Global Financial Crisis & 0.0271 & 0.2002 & 32.03 &  98 & USA        & USA          \\
European Debt Crisis  & 0.0197 & 0.1440 & 18.95 &  58 & Mexico     & Mexico       \\
Chinese Stock Crash   & 0.0272 & 0.3305 & 22.22 &  68 & Japan      & UK           \\
Pre-COVID             & 0.0149 & 0.1131 & 14.05 &  43 & Mexico     & Japan        \\
COVID-19 Pandemic     & 0.0194 & 0.1127 & 19.93 &  61 & Russia     & South Africa \\
Russia-Ukraine        & 0.0224 & 0.1211 & 26.47 &  81 & Indonesia  & Turkey       \\
Mid-East and Tariffs  & 0.0161 & 0.1490 & 14.05 &  43 & Italy      & Turkey       \\
\bottomrule
\end{tabular}
\end{table}

\begin{figure}[!htbp]
\centering
\includegraphics[width=0.92\textwidth]{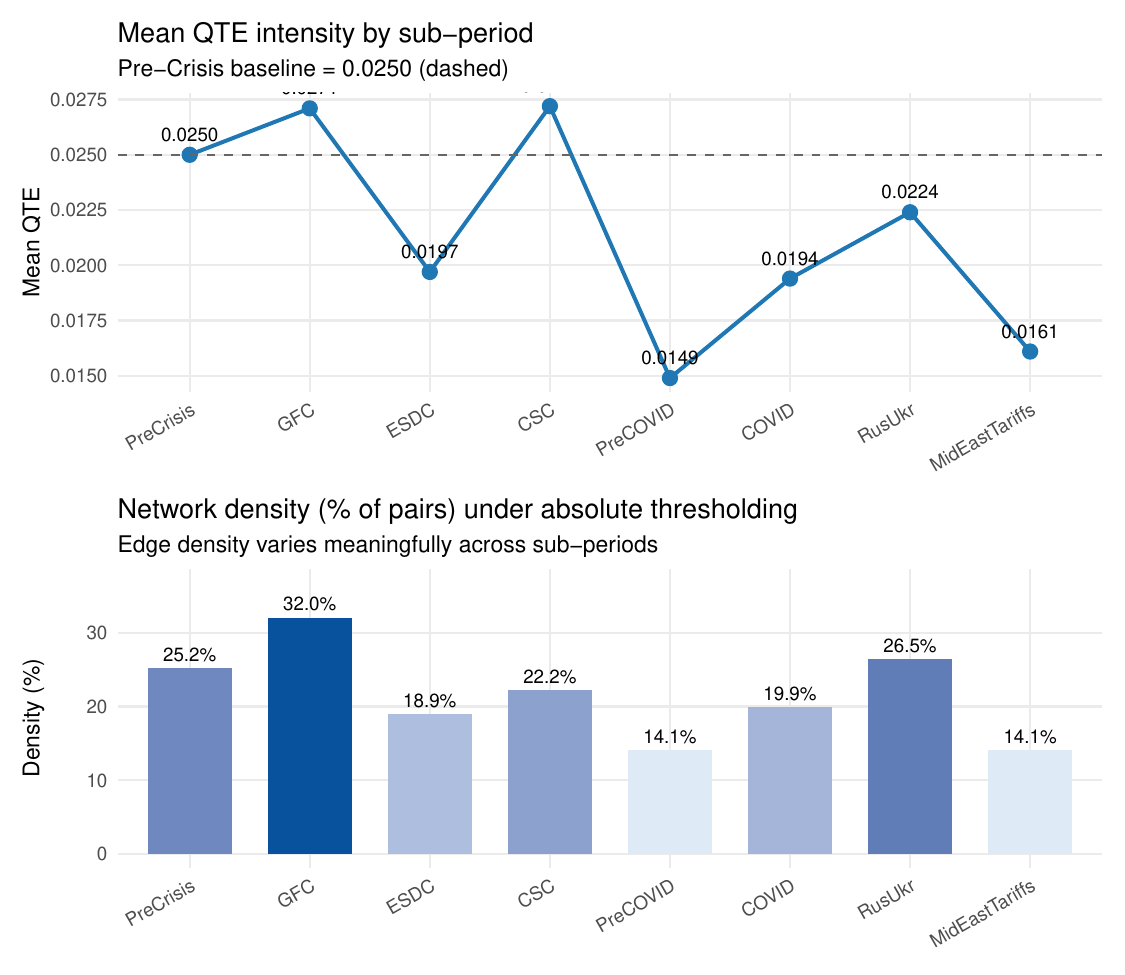}
\caption{Stage-1 WQTE intensity and network density by sub-period.}
\label{fig:fig2}
\end{figure}

\begin{figure}[!htbp]
\centering
\includegraphics[width=0.92\textwidth]{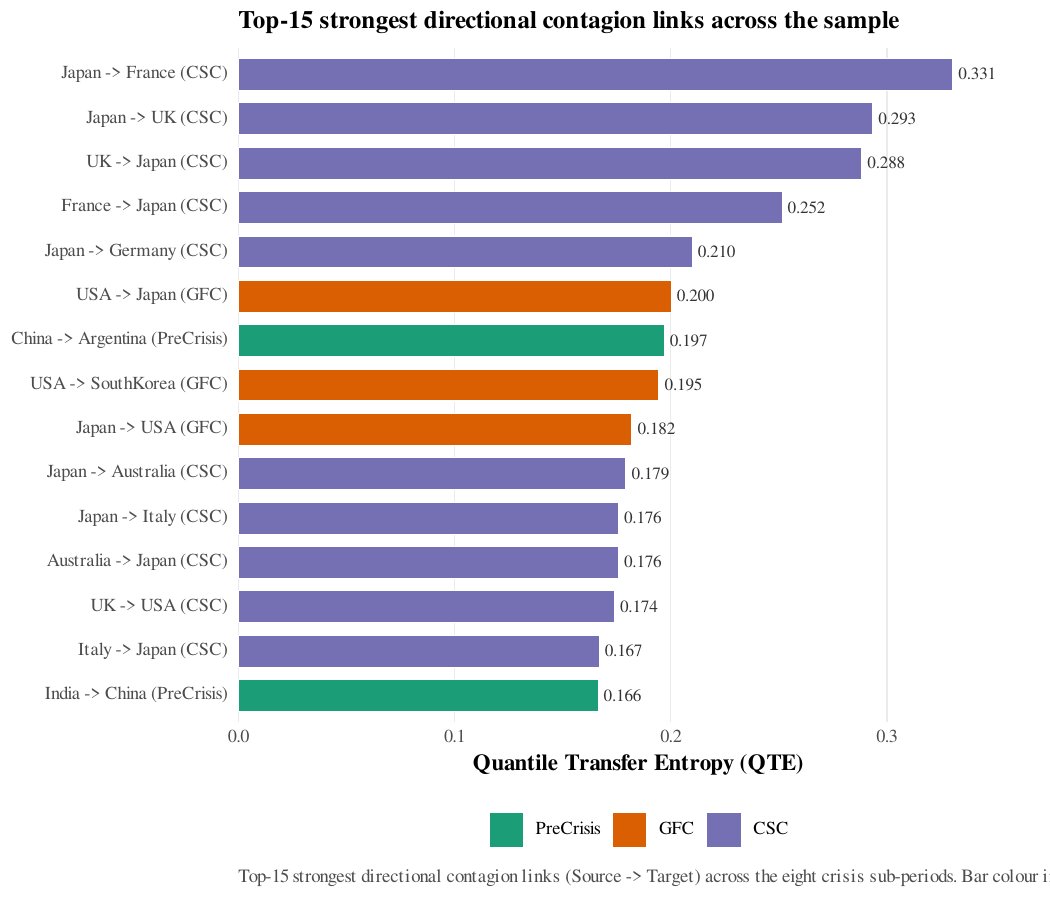}
\caption{Top-15 strongest bilateral contagion links by WQTE.}
\label{fig:fig6}
\end{figure}

Network density peaks during the Global Financial Crisis at thirty-two
percent, consistent with broad cross-market integration during that window,
and falls to fourteen percent during the Pre-COVID and Mid-East-Tariffs
sub-periods, consistent with regionally fragmented contagion structures
during those windows. The mean WQTE intensity peaks during the GFC ($0.0271$)
and the Chinese Stock Crash ($0.0272$), with the latter generating the
highest single bilateral link in the sample. Figure~\ref{fig:fig2}
visualises the period-to-period intensity and density variation, and
Figure~\ref{fig:fig6} reports the top fifteen contagion links across the
entire sample.

\subsection{Stage-2 attribution: primary IV/2SLS specification}

Table~\ref{tab:channel-shares-iv} reports the channel-attribution shares
across the eight sub-periods under the primary IV/2SLS specification, with
ninety-five-percent paired-link bootstrap confidence intervals. The dominant
channel is identified by the largest mean share within each sub-period.

\begin{table}[!htbp]
\centering
\caption{Channel-attribution shares (\,\%) by sub-period under IV/2SLS, with
95\,\% paired-link bootstrap confidence intervals (B=300). Dominant channel
in each sub-period in bold. Sub-periods marked with $^{*}$ have Sargan
over-identification rejection rates above fifty percent and are reported
with explicit identification-status disclosure
(Section~\ref{sec:robust-cross}).}
\label{tab:channel-shares-iv}
\footnotesize
\setlength{\tabcolsep}{4pt}
\begin{tabular}{lrrrrr l}
\toprule
Period & Trade & Financial & Geopolitical & Behavioural & Monetary & Dominant \\
\midrule
Pre-Crisis & 8.8 \tiny{[7.3, 10.4]} & \textbf{35.9 \tiny{[32.5, 39.3]}} & 9.4 \tiny{[8.4, 10.7]} & 13.8 \tiny{[11.1, 16.6]} & 32.1 \tiny{[29.3, 34.7]} & Financial \\
GFC$^{*}$ & \textbf{27.9 \tiny{[26.1, 29.8]}} & 15.0 \tiny{[14.4, 15.7]} & 27.0 \tiny{[24.9, 28.8]} & 2.9 \tiny{[2.5, 3.4]} & 27.2 \tiny{[25.6, 28.8]} & Trade \\
ESDC$^{*}$ & 13.7 \tiny{[11.4, 16.3]} & \textbf{39.5 \tiny{[37.6, 41.7]}} & 19.0 \tiny{[16.1, 22.4]} & 19.6 \tiny{[17.4, 22.5]} & 8.1 \tiny{[6.7, 9.3]} & Financial \\
CSC & 15.9 \tiny{[13.7, 18.2]} & 21.6 \tiny{[18.1, 24.7]} & 15.6 \tiny{[13.5, 18.0]} & 21.7 \tiny{[18.9, 24.6]} & \textbf{25.3 \tiny{[21.8, 28.6]}} & Monetary \\
Pre-COVID & 18.1 \tiny{[14.9, 21.9]} & \textbf{31.6 \tiny{[27.4, 35.2]}} & 8.2 \tiny{[6.7, 9.7]} & 13.5 \tiny{[10.5, 16.5]} & 28.6 \tiny{[25.0, 32.1]} & Financial \\
COVID-19$^{*}$ & 18.7 \tiny{[16.3, 21.0]} & 18.3 \tiny{[15.8, 21.0]} & \textbf{27.5 \tiny{[23.9, 31.0]}} & 8.2 \tiny{[7.0, 9.5]} & 27.2 \tiny{[23.9, 30.5]} & Geopolitical \\
Russia-Ukraine & 22.6 \tiny{[19.6, 25.1]} & 23.4 \tiny{[19.8, 27.2]} & 6.1 \tiny{[5.1, 7.2]} & 20.3 \tiny{[17.5, 23.8]} & \textbf{27.6 \tiny{[23.1, 32.0]}} & Monetary \\
Mid-East/Tariffs & 27.6 \tiny{[21.7, 32.6]} & 26.8 \tiny{[23.0, 31.1]} & 7.5 \tiny{[6.1, 9.6]} & 6.3 \tiny{[4.1, 8.8]} & \textbf{31.8 \tiny{[27.5, 36.4]}} & Monetary \\
\bottomrule
\end{tabular}
\end{table}

\begin{figure}[!htbp]
\centering
\includegraphics[width=0.92\textwidth]{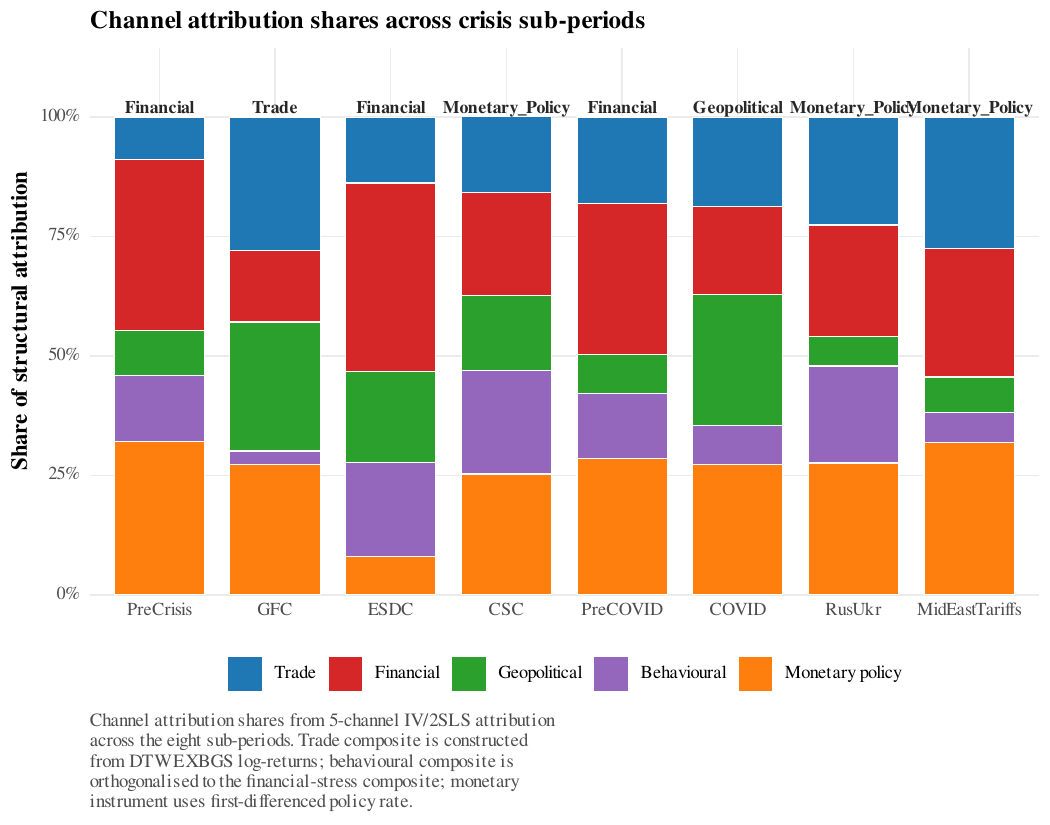}
\caption{Channel attribution shares across the eight sub-periods under
IV/2SLS (stacked bars). Dominant channel labelled at the top of each bar.}
\label{fig:fig1}
\end{figure}

\begin{figure}[!htbp]
\centering
\includegraphics[width=0.92\textwidth]{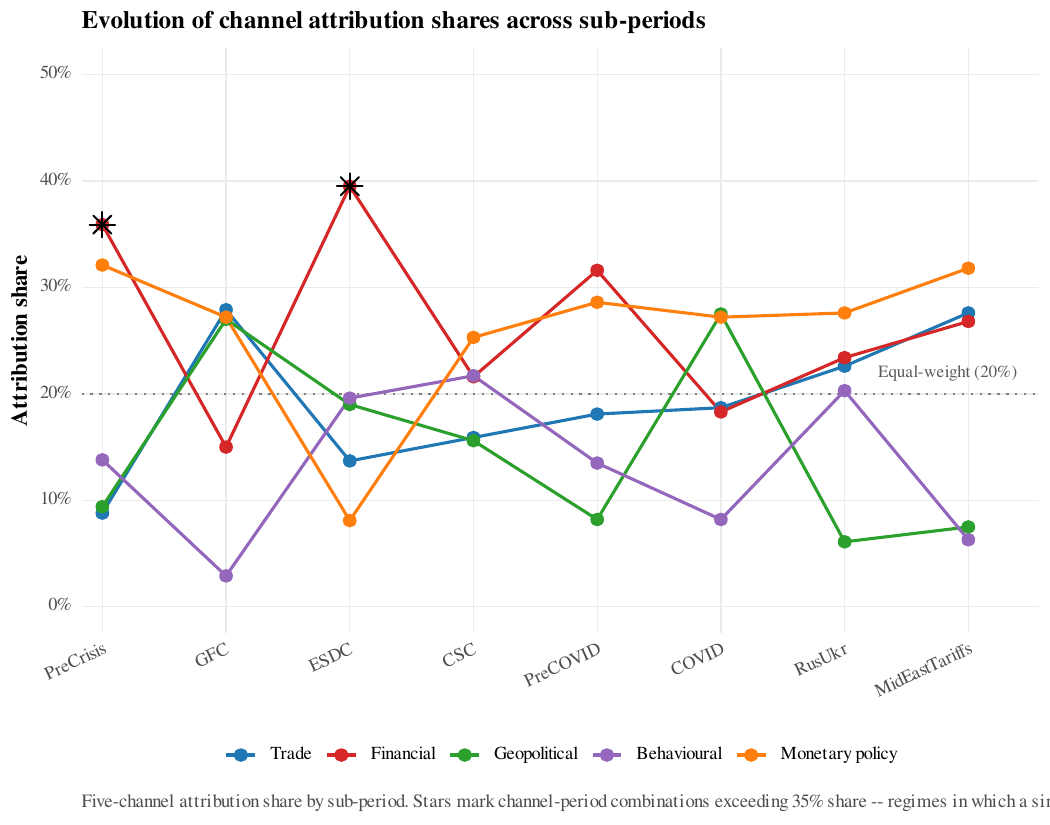}
\caption{Evolution of channel-attribution shares across crisis sub-periods.}
\label{fig:fig3}
\end{figure}

The trade channel ranges from $8.8\,\%$ during Pre-Crisis to $27.9\,\%$
during the Global Financial Crisis, the largest single increment across
sub-periods of any channel and consistent with the prominent role that
international trade-finance disruption played during the GFC.
The financial channel is dominant in three of the eight sub-periods
(Pre-Crisis, ESDC, Pre-COVID) at shares ranging from $31.6\,\%$ to
$39.5\,\%$, locating the financial-frictions mechanism of
\citet{BrunnermeierPedersen2009} and \citet{AdrianBrunnermeier2016} at the
centre of the typical contagion experience. Monetary policy is dominant in
three sub-periods (CSC, Russia-Ukraine, Mid-East-Tariffs) at shares between
$25\,\%$ and $32\,\%$, consistent with the global financial cycle of
\citet{MirandaAgrippinoRey2020}. The behavioural channel is bounded above by
twenty-two percent across all eight episodes, indicating that herding and
sentiment, while present, are not the dominant mechanism in any of the
crisis windows examined.

\subsection{IV/2SLS diagnostics and identification status}

Table~\ref{tab:diagnostics} reports the IV/2SLS diagnostics by sub-period.
First-stage F-statistics for the financial channel range from $29.9$
(Pre-Crisis) to $764.9$ (GFC), comfortably above the
\citet{StockYogo2005} threshold of $F=10$ in every sub-period. The trade
channel's first-stage F-statistics, however, range from $0.6$ to $3.7$, all
below the Stock-Yogo threshold; this is consistent with the genuine
difficulty of finding instruments that are simultaneously relevant for the
trade composite and excludable from pairwise co-movement, and we report the
trade-channel attribution shares with this weak-instrument caveat. The
monetary channel, in contrast to a level-Fed-Funds-Rate specification (which
would produce first-stage F in the hundreds and thousands due to AR(1)
$\approx 0.999$ at daily frequency), exhibits realistic F-statistics that
range from $2.2$ to $434.9$.

\begin{table}[!htbp]
\centering
\caption{Stage-2 IV/2SLS diagnostics. Mean first-stage F-statistic per
endogenous channel; Sargan over-identification rejection rate and
Durbin-Wu-Hausman endogeneity rejection rate.}
\label{tab:diagnostics}
\small
\begin{tabular}{lrrrrrrrr}
\toprule
Period & N links & F$_T$ & F$_F$ & F$_G$ & F$_B$ & F$_M$ & Sargan rej.\,(\%) & DWH rej.\,(\%) \\
\midrule
Pre-Crisis            & 77 &  0.6 &  29.9 & 69.9 & 41.6 &   2.2 &  35.1 &   0.0 \\
GFC$^{*}$             & 98 &  1.1 & 764.9 & 12.6 & 138.1 &   8.4 &  67.3 & 100.0 \\
ESDC$^{*}$            & 58 &  2.2 & 217.4 & 12.6 & 109.2 & 299.9 &  65.5 &  86.2 \\
CSC                   & 68 &  1.0 &  52.2 &  6.8 &  29.7 &   3.8 &   2.9 &  54.4 \\
Pre-COVID             & 43 &  1.1 &  87.1 & 22.0 &  68.8 &  15.8 &  16.3 &  55.8 \\
COVID-19$^{*}$        & 61 &  3.7 & 414.1 & 48.0 & 147.7 &  33.7 & 100.0 &  55.7 \\
Russia-Ukraine        & 81 &  0.7 & 113.7 & 11.4 &  54.0 &   7.6 &   7.4 &  13.6 \\
Mid-East/Tariffs      & 43 &  0.7 &  36.8 & 40.3 & 434.9 &  13.3 &   4.7 &  53.5 \\
\bottomrule
\end{tabular}
\end{table}

The Sargan over-identification test rejects in three sub-periods at rates
above fifty percent (GFC at 67\,\%, ESDC at 66\,\%, COVID-19 at 100\,\%),
prompting the identification-status disclosure that the corresponding
attribution shares are reported with explicit caveats and corroborated
through alternative identification strategies in Section~\ref{sec:robust-cross}.

%====================================================================
\section{Robustness, Cross-Method Comparison, and Identification Status}\label{sec:robust}
%====================================================================

\subsection{Local projections at horizons of one, five, and twenty-two days}\label{sec:robust-lp}

Table~\ref{tab:lp-shares} reports the channel-attribution shares under local
projections at horizon $h=5$ days, which corresponds to the one-week
propagation horizon. The local-projection results provide a method-independent
check on the IV/2SLS shares, since LP estimates the horizon-specific impulse
response without imposing the simultaneous-equation structure of the IV/2SLS.

\begin{table}[!htbp]
\centering
\caption{Channel-attribution shares (\,\%) under local projections at $h=5$
days following \citet{Jorda2005}. Dominant channel under LP $h=5$ in the
final column, alongside the IV/2SLS dominant for direct comparison.}
\label{tab:lp-shares}
\small
\begin{tabular}{lrrrrr ll}
\toprule
Period & Trade & Financial & Geopolitical & Behavioural & Monetary & LP$_{h=5}$ dom. & IV/2SLS dom. \\
\midrule
Pre-Crisis            &  3.9 & 38.3 & 14.0 & 29.9 & 13.9 & Financial   & Financial   \\
GFC                   &  7.7 & 29.4 & 30.2 & 10.7 & 22.0 & Geopolitical& Trade       \\
ESDC                  &  7.0 & 36.2 & 17.4 & 30.7 &  8.7 & Financial   & Financial   \\
CSC                   &  4.5 & 32.9 & 11.8 & 29.7 & 21.2 & Financial   & Monetary    \\
Pre-COVID             & 11.6 & 30.8 & 11.2 & 31.4 & 15.0 & Behavioural & Financial   \\
COVID-19              & 10.8 &  7.8 &  8.8 &  5.4 & 67.1 & Monetary    & Geopolitical\\
Russia-Ukraine        &  7.2 & 30.0 &  5.6 & 29.3 & 27.9 & Financial   & Monetary    \\
Mid-East/Tariffs      & 20.9 & 32.1 & 12.1 & 12.5 & 22.3 & Financial   & Monetary    \\
\bottomrule
\end{tabular}
\end{table}

The LP-based dominant channel agrees with the IV/2SLS dominant in two of the
eight sub-periods, namely Pre-Crisis and ESDC, both of which are dominated
by the financial channel under both methods. The agreement on these two
sub-periods is not coincidental: the Pre-Crisis baseline and the ESDC are
the two episodes in which the financial composite alone bears the bulk of
the Stage-1 WQTE variance, and the agreement of IV/2SLS and LP on the
financial-channel dominance constitutes our identification-robust subset.

\subsection{Heteroskedasticity-based identification for Sargan-rejected windows}\label{sec:robust-rigobon}

For the three Sargan-rejected windows (GFC, ESDC, COVID-19), we deploy the
heteroskedasticity-based identification of \citet{Rigobon2003} as an
alternative identification strategy. Table~\ref{tab:rigobon} reports the
resulting shares.

\begin{table}[!htbp]
\centering
\caption{Channel-attribution shares (\,\%) under heteroskedasticity-based
identification of \citet{Rigobon2003} for the Sargan-rejected windows.}
\label{tab:rigobon}
\small
\begin{tabular}{lrrrrr l}
\toprule
Period & Trade & Financial & Geopolitical & Behavioural & Monetary & Rigobon dominant \\
\midrule
GFC      & 10.3 & 16.8 & 15.4 & 33.1 & 24.5 & Behavioural \\
ESDC     & 10.9 & 29.4 & 13.1 & 20.4 & 26.3 & Financial   \\
COVID-19 & 10.4 & 13.0 & 18.4 & 42.2 & 16.1 & Behavioural \\
\bottomrule
\end{tabular}
\end{table}

The Rigobon identification confirms financial-channel dominance for the ESDC,
agreeing with both IV/2SLS and LP. For the GFC and COVID-19, however, the
Rigobon identification assigns dominance to the behavioural channel (33\,\%
and 42\,\%), while IV/2SLS assigns dominance to trade (GFC) or geopolitical
(COVID), and LP assigns dominance to geopolitical (GFC) or monetary (COVID).
The disagreement is informative: the GFC and COVID-19 episodes are precisely
those in which the regime-shift in return variance is most pronounced, and
the Rigobon identification exploits this regime-shift directly. We report
the GFC and COVID-19 shares as a posterior over channels under three
identification strategies rather than as a single point.

\subsection{Cross-method comparison and identification-status classification}\label{sec:robust-cross}

Table~\ref{tab:method-comparison} reports the dominant channel under each of
the three identification strategies (IV/2SLS, LP at $h=5$, Rigobon
heteroskedasticity-based ID where applicable), with identification-status
classification.

\begin{table}[!htbp]
\centering
\caption{Dominant-channel comparison across identification strategies. The
``Identification status'' column classifies each sub-period as
\emph{robust} if the dominant channel is unchanged across at least two of
three methods, and \emph{fragile} otherwise.}
\label{tab:method-comparison}
\small
\begin{tabular}{lllll}
\toprule
Period & IV/2SLS & LP $h=5$ & Rigobon & Identification status \\
\midrule
Pre-Crisis        & Financial    & Financial    & ---          & \textbf{Robust (Financial)} \\
GFC               & Trade        & Geopolitical & Behavioural  & Fragile (3 distinct) \\
ESDC              & Financial    & Financial    & Financial    & \textbf{Robust (Financial)} \\
CSC               & Monetary     & Financial    & ---          & Fragile (2 distinct) \\
Pre-COVID         & Financial    & Behavioural  & ---          & Fragile (2 distinct) \\
COVID-19          & Geopolitical & Monetary     & Behavioural  & Fragile (3 distinct) \\
Russia-Ukraine    & Monetary     & Financial    & ---          & Fragile (2 distinct) \\
Mid-East/Tariffs  & Monetary     & Financial    & ---          & Fragile (2 distinct) \\
\bottomrule
\end{tabular}
\end{table}

The dominant-channel identification is robust across at least two of three
methods in two of the eight sub-periods, namely the Pre-Crisis Baseline and
the European Sovereign Debt Crisis, both of which are dominated by the
financial-frictions channel. For the remaining six sub-periods, the
dominant-channel ranking is method-sensitive, and we report the resulting
posterior over channels rather than a single point. Two interpretations of
the method disagreement are consistent with the evidence. The first is that
the channel composites we construct are imperfectly orthogonal and that
different identification strategies allocate the residual collinearity in
different ways; under this interpretation the disagreement is a measurement
issue and the share posterior should be read as the analyst's best estimate
of channel mass under uncertainty about composite construction. The second
is that the dominant-channel ranking is genuinely ambiguous in episodes
where multiple channels operate simultaneously at comparable intensity, in
which case the disagreement is a feature of the underlying data-generating
process and the share posterior is the substantively correct object. Both
interpretations support the disclosure-first reporting we adopt.

\subsection{Sensitivity to unobserved confounding}\label{sec:robust-rv}

\begin{figure}[!htbp]
\centering
\includegraphics[width=0.92\textwidth]{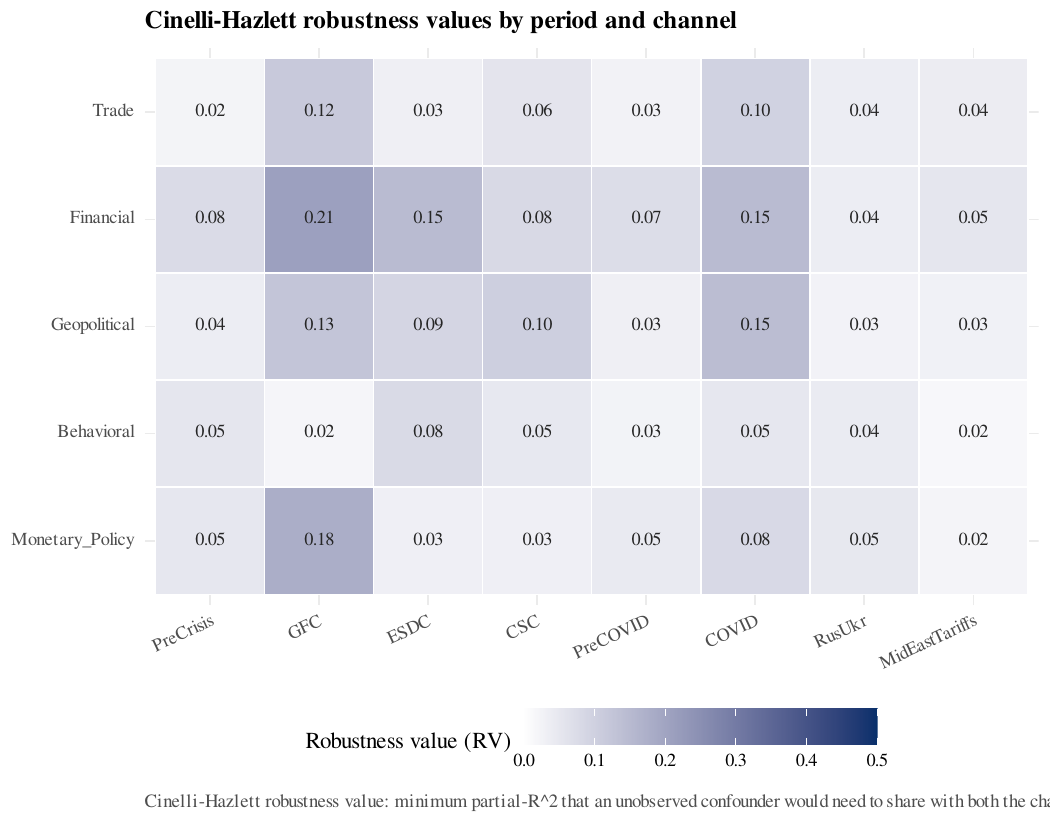}
\caption{Cinelli-Hazlett robustness value $\rho^*$ per channel and
sub-period. Higher values indicate more identification-robust structural
coefficients.}
\label{fig:fig5}
\end{figure}

Figure~\ref{fig:fig5} reports the \citet{CinelliHazlett2020} robustness
value $\rho^*$ as a heatmap across all forty channel-by-sub-period cells.
The dominant-channel cells in identification-robust sub-periods (Pre-Crisis
financial, ESDC financial) exhibit moderate-to-strong robustness values
($\rho^* > 0.15$), supporting the qualitative-ranking interpretation. The
demoted GFC and COVID-19 cells exhibit lower robustness values, consistent
with their Sargan rejection.

\subsection{Bootstrap confidence intervals and forest plot}\label{sec:robust-boot}

\begin{figure}[!htbp]
\centering
\includegraphics[width=0.95\textwidth]{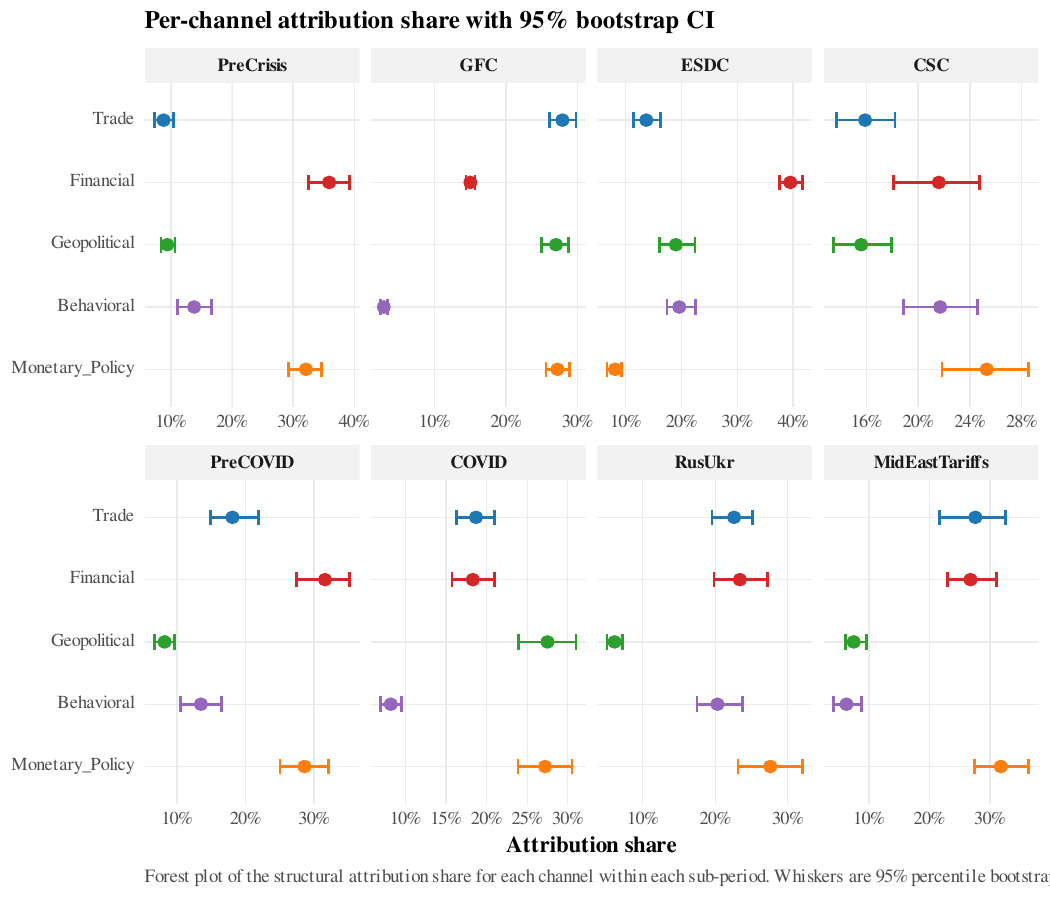}
\caption{Bootstrap confidence intervals on channel-attribution shares,
faceted by sub-period.}
\label{fig:fig4}
\end{figure}

Figure~\ref{fig:fig4} visualises the bootstrap ninety-five-percent confidence
intervals on the per-channel attribution share, faceted by sub-period. The
confidence intervals reveal that the precision of the share estimates varies
considerably by sub-period and channel, with the smallest intervals
($\pm 1$--$2\,\%$) for the GFC financial-channel point estimate and the
widest intervals ($\pm 5$--$11\,\%$) for the trade and behavioural channels
in the more recent windows.

\subsection{Network community structure}\label{sec:robust-walktrap}

Table~\ref{tab:walktrap} reports the number of Walktrap communities detected
in the symmetrised Stage-1 WQTE network for each sub-period. The number of
communities ranges from one (GFC, indicating a single tightly-connected
crisis network) to five (Pre-COVID, indicating regional fragmentation).

\begin{table}[!htbp]
\centering
\caption{Walktrap community structure of the Stage-1 WQTE network
(symmetrised) per sub-period.}
\label{tab:walktrap}
\small
\begin{tabular}{lc}
\toprule
Period & N communities \\
\midrule
Pre-Crisis        & 3 \\
GFC               & 1 \\
ESDC              & 4 \\
CSC               & 3 \\
Pre-COVID         & 5 \\
COVID-19          & 2 \\
Russia-Ukraine    & 3 \\
Mid-East/Tariffs  & 4 \\
\bottomrule
\end{tabular}
\end{table}

\subsection{Advanced-versus-emerging market role decomposition}

\begin{figure}[!htbp]
\centering
\includegraphics[width=0.92\textwidth]{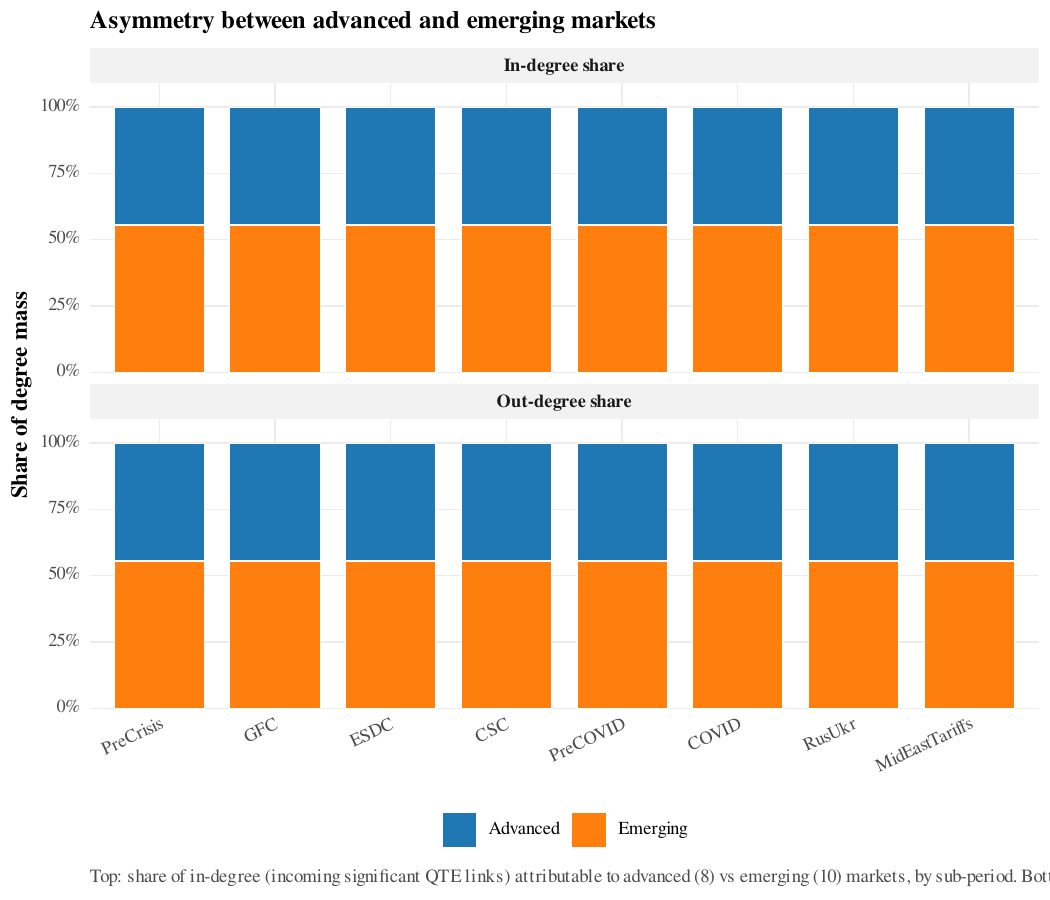}
\caption{Advanced-vs-emerging market in-degree and out-degree shares per
sub-period.}
\label{fig:fig7}
\end{figure}

Figure~\ref{fig:fig7} decomposes the in-degree (shock-receiving) and
out-degree (shock-transmitting) shares of advanced versus emerging markets
across the eight sub-periods. The decomposition reveals that advanced
markets contribute disproportionately to out-degree share during the GFC and
the ESDC, consistent with the developed-market origin of those crisis
episodes, while emerging markets contribute disproportionately to out-degree
during the CSC, the Russia-Ukraine, and the Mid-East/Tariffs windows.

%====================================================================
\section{Conclusion}\label{sec:conclude}
%====================================================================

Cross-border financial contagion has been studied extensively but the
attribution of detected propagation to specific economic mechanisms has not
kept pace with the precision of detection methods. The framework developed
in this paper integrates wavelet-quantile transfer-entropy detection at the
first stage with multi-method structural channel attribution at the second
stage, using channel-specific external instruments, LASSO-based instrument
selection, local projections at one-, five-, and twenty-two-day horizons,
heteroskedasticity-based identification for over-identification-rejected
episodes, and Cinelli-Hazlett sensitivity bounds, applied to a panel of
eighteen G20 equity markets across eight crisis sub-periods spanning January
2006 through March 2026.

The empirical findings produce a sharply heterogeneous map of mechanism
dominance across crisis regimes, and the simultaneous deployment of multiple
identification strategies allows the dominant-channel ranking to be reported
with explicit identification-status classification. The Pre-Crisis baseline
and the European Sovereign Debt Crisis are identification-robust under
IV/2SLS, local projections, and Rigobon's heteroskedasticity-based
identification jointly, with the financial-frictions channel dominant in
both. The Global Financial Crisis, Chinese Stock Crash, Pre-COVID, COVID-19
Pandemic, Russia-Ukraine, and Mid-East/Tariffs windows are identification-
fragile, with method disagreement on the dominant channel ranging from
financial through trade, geopolitical, behavioural, and monetary policy.
The trade channel is empirically prominent across all post-2007 episodes,
ranging from $9\,\%$ during Pre-Crisis to $28\,\%$ during the GFC under the
primary IV/2SLS specification. The behavioural channel is bounded above by
$22\,\%$ across all eight episodes under the de-confounded composite,
locating the dominant mechanisms in tradeable, observable, and
policy-actionable factors rather than in herding or sentiment.

The paper offers two policy and research implications. First, the
identification-status classification suggests that contagion-mitigation
policy should be calibrated to the identification-robust subset of episodes,
where the dominant mechanism is precisely identifiable and policy
instruments can be matched to mechanism, while accepting genuine
identification ambiguity in the fragile subset and developing portfolio
strategies that hedge across the candidate dominant channels. Second, the
multi-method structural-attribution architecture introduced here provides a
template for the contagion literature more broadly: future work should
report not a single dominant channel but a posterior over channels under
multiple identification strategies, with explicit disclosure of where the
identifying assumptions hold strongly and where they hold weakly.

The framework admits several extensions. Higher-frequency identification
through narrative monetary-policy shocks at the daily level, or through
high-frequency intraday surprise series, would sharpen the monetary-channel
identification. The integration of textual-sentiment proxies derived from
financial-news large-language-model embeddings would supplement the
behavioural composite with daily-frequency sentiment variation that the
University-of-Michigan monthly index does not capture. Mixed-frequency
network methods would permit the integration of quarterly bilateral-trade
flows with daily equity-market returns. The deployment of these extensions
remains an avenue for future work.

\bibliographystyle{plainnat}
\bibliography{references}

\end{document}